\documentclass[
    ,final            % use final for the camera ready runs
  ]
  {aipproc}

\layoutstyle{8x11single}
\usepackage{amsmath,amssymb,bm}

\begin{document}

\title{Photoproduction of $\bm{K} \bm{\Sigma(1385)}$ from the nucleon}


\author{Yongseok Oh}{
address={School of Physics and Energy Sciences,
	Kyungpook National University,
	Daegu 702-701, Korea}
}

\begin{abstract}
The reactions of $K \Sigma(1385)$ photoproduction, i.e., $\gamma p \to K^+
\Sigma^0(1385)$ and $\gamma n \to K^+ \Sigma^-(1385)$, are investigated in the 
resonance energy region for studying the role of the nucleon and $\Delta$ 
resonances of masses around 2 GeV.
The Lagrangians for describing the decays of these resonances into the
$K \Sigma(1385)$ channel are constructed and the decay amplitudes are
obtained, which allows us to determine the coupling constants using the
predictions of quark models or the data listed by the 
Particle Data Group. 
The resulting cross sections are compared to the data from the Thomas
Jefferson National Accelerator Facility and the SPring-8, which indicates
nontrivial contributions from the two-star-rated resonances in the
Particle Data Group as well as from some missing resonances predicted by
a quark model. 
\end{abstract}

\maketitle

%%%%%%%%%%%%%%%%%%%%%%%%%%%%%%%%%%%%%%%%%%%%
%% MAINMATTER
%%%%%%%%%%%%%%%%%%%%%%%%%%%%%%%%%%%%%%%%%%%%

Production of strange particles from photon-nucleon scattering has been
extensively studied in recent experiments at electron/photon
accelerator facilities~\cite{CLAS05c,LEPS03b,SAPHIR98,TTWF07}.
However, most of the data accumulated by these experiments are for
the reactions of $K\Lambda(1116)$ and $K\Sigma(1193)$ production,
and the photoproduction of the spin-$3/2$ hyperons such as the
reaction of $\gamma p \to K \Sigma(1385)$ is virtually unknown until the
recent experiments at JLab~\cite{CLAS06e} and at SPring-8~\cite{LEPS08b}.
The $\Sigma^*$ or $\Sigma(1385)$ is the lowest mass hyperon in the baryon
decuplet, so the analysis of its production mechanism will be valuable for
testing the flavor SU(3) symmetry when combined with the analyzed production
mechanism of $\Delta(1232)$ resonance.

Furthermore, this reaction is expected to have a nontrivial role in
searching for the so-called missing resonances.
This is based on the observation that, although the cross section of
$\Sigma(1385)$ photoproduction is smaller than that of $K\Lambda(1116)$
and $K\Sigma(1193)$ photoproduction, the suppression factor is not large,
which means that this reaction may have nontrivial role in a full coupled
channel calculation for identifying new resonances and/or extracting the
properties of baryon resonances.%
\footnote{Note that similar conclusions were drawn also for $K^* \Lambda$
and $K^* \Sigma$ photoproduction~\cite{CLAS06e,CLAS07a,CBELSA-08,OK06}.}
In fact, the quark model of Ref.~\cite{CR98b} predicts that several
missing and not-well-established non-strange baryon resonances have large
partial decay widths into the $K\Sigma(1385)$ channel, while the couplings
of most well-established resonances to this channel are small.
Then we may have a chance to see the effects of such missing or
yet-to-be-established resonances
through the analysis of $K\Sigma(1385)$ photoproduction process.

The recent experiments for $K\Sigma(1385)$ photoproduction were reported
in Refs.~\cite{CLAS06e,LEPS08b}, which present the cross sections and the
beam asymmetry of this reaction.%
\footnote{The earlier experimental data for $K\Sigma(1385)$
photoproduction obtained in 1960's can be found, e.g., in
Refs.~\cite{CBCG67,DBCG67}.}
Theoretical investigation of $K\Sigma(1385)$ photoproduction, however, is
rare. (See, for example, Refs.~\cite{LS05,DOS06}.)
Here we present the theoretical model of Ref.~\cite{OKN08} for the
reaction mechanisms of $\Sigma(1385)$ production in the reaction of
$\gamma p \to K^+ \Sigma^0(1385)$, which is based on the preliminary
data of the JLab~\cite{CLAS06e}.
We then discuss the application of this model to the physical
quantities in the reaction of $\gamma n \to K^+ \Sigma^-(1385)$, of which
data were reported in Ref.~\cite{LEPS08b}.

\begin{figure}[t]
\includegraphics[width=.9\textwidth]{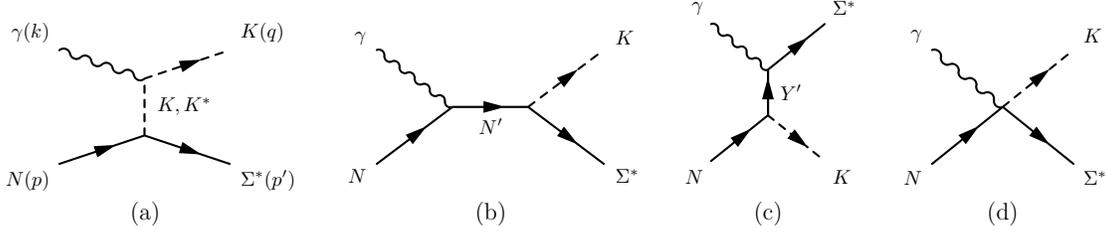}
\caption{Feynman diagrams for $K\Sigma^*$ photoproduction.
$N'$ stands for the non-strange baryons and their resonances, while $Y'$ the
hyperons with strangeness $-1$ and their resonances.
Diagram (d) includes the contact term of which form can be found in
Ref.~\cite{OKN08}.}
\label{fig:diag}
\end{figure}

Our model for $K\Sigma^*$ photoproduction is depicted in
Fig.~\ref{fig:diag}.
This includes the $t$-channel $K$ and $K^*$ meson exchanges
[Fig.~\ref{fig:diag}(a)], $s$-channel non-strange baryon exchanges
[Fig.~\ref{fig:diag}(b)], and $u$-channel hyperon exchanges
[Fig.~\ref{fig:diag}(c)].
It also contains the generalized contact term [Fig.~\ref{fig:diag}(d)]
that is introduced to restore gauge invariance.
For the effective Lagrangians to calculate the production
amplitudes from the diagrams for the $t$-channel $K$ exchange,
$s$-channel nucleon term, $u$-channel $\Sigma(1385)$ term, and the
contact term, we refer the reader to Ref.~\cite{OKN08}, where the details
of the effective Lagrangians and the coupling constants are discussed.
Instead, we here focus on the Lagrangians which require more rigorous and
detailed studies.

The $t$-channel $K^*$ exchange involves the $\gamma K K^*$ and $K^* N
\Sigma^*$ interactions.
The Lagrangian for the former interaction is well known and its coupling
constant can be determined from the measured width of the $K^* \to K
\gamma$ decay. 
However, the Lagrangian for the latter interaction, which describes the
vertex of $\frac32 \to 1 + \frac12$ with the numbers representing the spin
of the participating particles, has not been studied in detail.
This interaction contains three terms in general, which can be written as
\begin{eqnarray}
\mathcal{L}_{K^*N\Sigma^*} &=& \frac{ig_1^{}}{2M_N}
\overline{K}^{*\mu\nu} \overline{\bm\Sigma}^*_\mu \cdot \bm{\tau}
\gamma_\nu \gamma_5^{} N
+ \frac{g_2^{}}{(2M_N)^2} \overline{K}^{*\mu\nu}
\overline{\bm\Sigma}^*_\mu \cdot \bm{\tau} \gamma_5^{} \partial_\nu N
- \frac{g_3^{}}{(2M_N)^2}
\partial_\nu \overline{K}^{*\mu\nu} \overline{\bm\Sigma}^*_\mu \cdot
\bm{\tau} \gamma_5^{} N + \mbox{ H.c.},
\end{eqnarray}
where $K^*_{\mu\nu} = \partial_\mu K^*_\nu - \partial_\nu K^*_\mu$.
The coupling constants of this interaction is unknown and may be
determined from the corresponding couplings of the $\rho N \Delta$
interaction if we invoke the SU(3) symmetry relation.
However, even the couplings of the $\rho N \Delta$ interaction are poorly
known and, in most analysis, only the $g_1^{}$ term has been
used~\cite{KA04,ONL04}.
In this work, we fix the coupling $g_1^{}$ of the $K^* N \Sigma^*$
interaction by using the SU(3) symmetry with that of the $\rho N \Delta$
interaction and treat $g_2^{}$ and $g_3^{}$ as free parameters.
Fortunately, it turns out that in our case, the contribution from the
$K^*$ exchange is suppressed and the uncertainties of those couplings
could be ignored. However, it would be worth while to analyze the data
for estimating all three couplings of the $\rho N \Delta$ interaction.

For the $u$-channel diagrams, we consider the case with $Y' =
\Lambda(1116)$ and $\Sigma(1385)$. The former is the lowest mass hyperon
and the latter is required to fulfill the gauge invariance condition.
Since one of the motivation of this study is to investigate the
non-strange baryon resonances in the $s$-channel diagrams, we start with
the most general expressions for the interactions of $R N \gamma$ and
$R K \Sigma^*$, which read
\begin{eqnarray}
\mathcal{L}_{RN\gamma}(\textstyle\frac12^\pm) &=& \frac{ef_1}{2M_N}
\bar{N} \Gamma^{(\mp)} \sigma_{\mu\nu} \partial^\nu A^\mu R +
\mbox{H.c.},
\nonumber \\
\mathcal{L}_{RN\gamma}(\textstyle\frac32^\pm) &=& -
\frac{ief_1}{2M_N} \overline{N} \Gamma^{(\pm)}_\nu F^{\mu\nu} R_\mu
- \frac{ef_2}{(2M_N)^2} \partial_\nu \bar{N}
\Gamma^{(\pm)} F^{\mu\nu} R_\mu + \mbox{H.c.},
\nonumber \\
\mathcal{L}_{RN\gamma}(\textstyle\frac52^\pm) &=&
\frac{ef_1}{(2M_N)^2} \bar{N} \Gamma_\nu^{(\mp)}
\partial^\alpha F^{\mu\nu} R_{\mu\alpha}
-\frac{ief_2}{(2M_N)^3} \partial_\nu \bar{N} \Gamma^{(\mp)}
\partial^\alpha F^{\mu\nu} R_{\mu\alpha}
+ \mbox{H.c.},
\nonumber \\
\label{eq:RNgamma}
\end{eqnarray}
and
\begin{eqnarray}
\mathcal{L}_{RK\Sigma^*}(\textstyle\frac12^\pm) &=& \frac{h_1}{M_K}
\partial_\mu K \bar{\Sigma}^{*\mu} \Gamma^{(\mp)} R + \mbox{H.c.},
\nonumber \\
\mathcal{L}_{RK\Sigma^*}(\textstyle\frac32^\pm) &=& \frac{h_1}{M_K}
\partial^\alpha K \bar{\Sigma}^{*\mu} \Gamma_\alpha^{(\pm)} R_\mu
+ \frac{ih_2}{M_K^2} \partial^\mu
\partial^\alpha K \bar{\Sigma}^*_\alpha \Gamma^{(\pm)} R_\mu +
\mbox{H.c.},
\nonumber \\
\mathcal{L}_{RK\Sigma^*}(\textstyle\frac52^\pm) &=& \frac{i h_1}{M_K^2}
\partial^\mu \partial^\beta K \bar{\Sigma}^{*\alpha} \Gamma_\mu^{(\mp)}
R_{\alpha\beta}
- \frac{h_2}{M_K^3}\partial^\mu
\partial^\alpha \partial^\beta K \bar{\Sigma}^*_\mu \Gamma^{(\mp)}
R_{\alpha\beta} + \mbox{H.c.}.
\nonumber \\
\label{eq:RKSigma*}
\end{eqnarray}
where $A_\mu$ is the photon field ($F_{\mu\nu} = \partial_\mu A_\nu -
\partial_\nu A_\mu$), and $R$, $R_\mu$, and $R_{\mu\nu}$ are the
fields for the spin-$1/2$, $3/2$, and $5/2$ resonances, respectively, with
\begin{equation}
\Gamma^{(\pm)}_\mu = \left( \begin{array}{c} \gamma_\mu \gamma_5^{} \\
\gamma_\mu \end{array} \right), \qquad \Gamma^{(\pm)} = \left(
\begin{array}{c} \gamma_5^{} \\
\textbf{1} \end{array} \right).
\label{gamma_pm}
\end{equation}
The coupling constants $f_i^{}$ and $h_i^{}$ can be related to the
theoretical predictions on the partial decay amplitudes of the resonances.
This has a great advantage compared with the use of partial decay widths
as the relative phase of the coupling constants can be fixed.
The explicit formulas for this relation were derived in
Ref.~\cite{OKN08}, and the couplings fixed by the quark model
prediction of Refs.~\cite{CR98b} are listed in
Table~\ref{tab:coup}.%
\footnote{The typographical errors committed in Ref.~\cite{OKN08} are
corrected here.}
In this work, only the resonances which are predicted to have large
couplings to the $K\Sigma^*$ channel in the model of Ref.~\cite{CR98b}
are considered.

\begin{table}[t]
\centering\small
\begin{tabular}{c|c|cc|cc}
\hline \textbf{Resonance} & \textbf{PDG}~\cite{PDG08} &
$h_1$ & $h_2$ & $f_1$ & $f_2$ \\ \hline
$N\frac12^-(1945)$ & $S_{11}^{*}(2090)$ & 
$-0.98$  & --- & $0.055$  & ---
\\
$N\frac32^-(1960)$ & $ D_{13}^{**}(2080)$ & 
$0.24$ & $-0.54$ & $-1.25$ & $1.21$
\\
$N\frac52^-(2095)$ & $ D_{15}^{**}(2200)$ &
$0.29$ & $-0.08$ & $0.37$ & $0.28$
\\
$\Delta\frac32^-(2080)$ & $ D_{33}^{*}(1940)$ &
$-0.68$ & $1.00$ & $0.39$ & $-0.57$
\\
$\Delta\frac52^+(1990)$ & $ F_{35}^{**}(2000)$ &
$-0.87$ & $0.11$ & $-0.68$ & $-0.062$
\\ \hline
$N\frac32^-(2095)$ & & $0.99$ & $0.27$ & $0.49$ & $-0.83$
\\
$N\frac52^+(1980)$ & & $0.59$ & $0.24$ & $0.019$ & $-0.13$
\\
$\Delta\frac32^-(2145)$ & & $0.25$ & $0.46$ & $0.11$ & $-0.059$ \\
\hline
\end{tabular}
\caption{Resonances and their coupling constants $h_i^{}$ and $f_i^{}$
based on the predictions of Ref.~\cite{CR98b}. The coupling
constants are calculated using the resonance masses of PDG.}
\label{tab:coup}
\end{table}

\begin{figure}[b]
\includegraphics[height=.3\textheight]{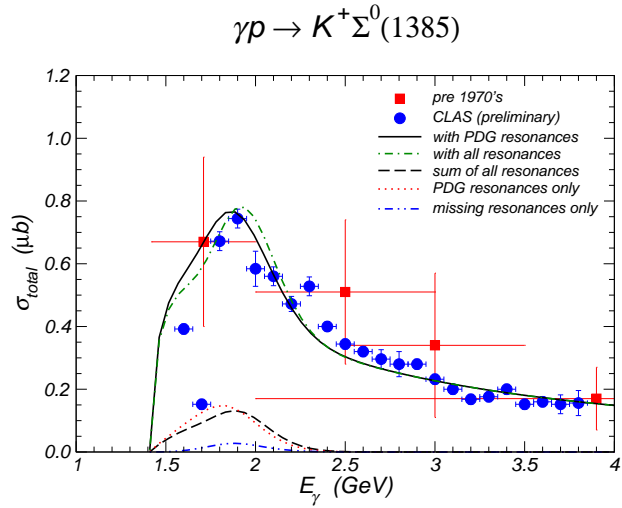}
\caption{Total cross sections for the $\gamma p \to
K^+ \Sigma^0(1385)$ reaction with the resonance parameters listed in
Table~\ref{tab:coup}.}
\label{fig:total-res}
\end{figure}

Presented in Fig.~\ref{fig:total-res} is the result of the total cross
section for the $\gamma p \to K^+ \Sigma^0(1385)$ reaction, which
indicates nontrivial contribution from the missing and/or
not-well-established resonances in production mechanism of this reaction.
In particular, we find that nontrivial contributions from the
$\Delta(2000)F_{35}$, $\Delta(1940)D_{33}$, and $N(2080)D_{13}$, as well as
from the missing resonance $N\frac{3}{2}^-(2095)$ are required to
describe the observed data of Ref.~\cite{CLAS06e}.

\begin{figure}[t]
\includegraphics[width=.4\textwidth]{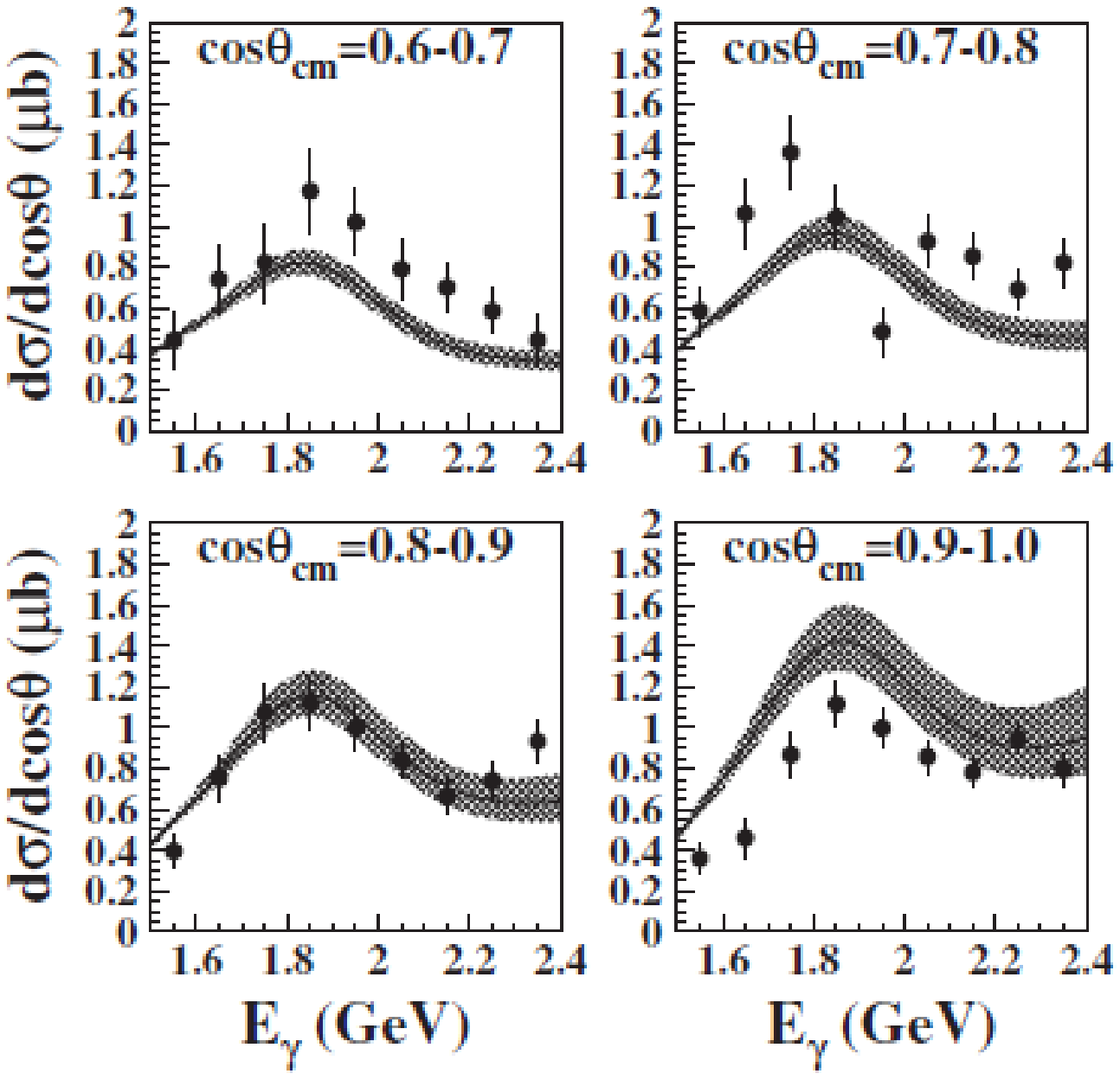}
\includegraphics[width=.4\textwidth]{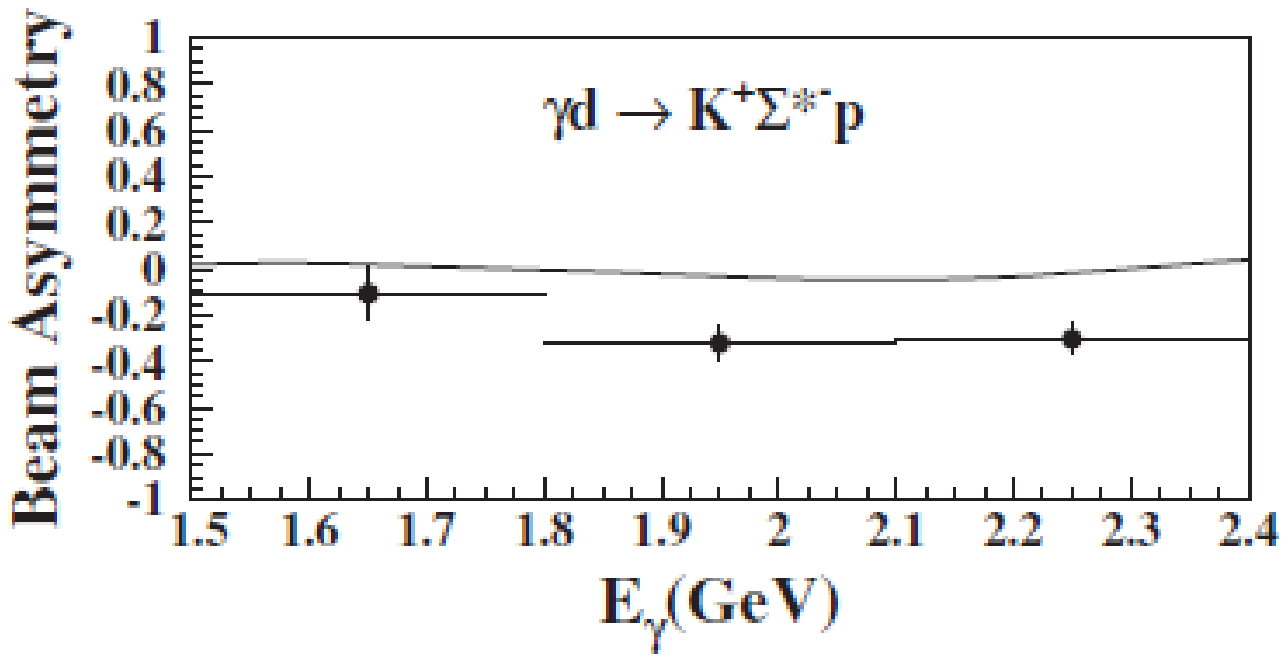}
\caption{(Left panel) Differential cross sections for the $\gamma n \to
K^+ \Sigma^-(1385)$ reaction with the resonances listed in
Table~\ref{tab:coup}. (Right panel) Beam asymmetry of the same reaction.}
\label{fig:diff-res}
\end{figure}

The same model is then applied to the reaction of $\gamma n \to
K^+ \Sigma^-(1385)$ with the same resonances and keeping the isospin
symmetry. The predictions of this model for differential cross sections
and beam asymmetry are compared with the data obtained by the LEPS
Collaboration~\cite{LEPS08b} in Fig.~\ref{fig:diff-res}.
Considering that only a few resonances are included and no additional
parameters are allowed, the description of this model for the differential
cross sections is reasonable.
However, the difficulty in explaining the beam asymmetry indicates that
more sophisticated and detailed analyses of the data are highly required
to understand the production mechanism and to pin down the role of each
baryon resonance.

In summary, we have explored the reactions of $K\Sigma(1385)$
photoproduction from the proton and the neutron targets.
We have constructed the Lagrangians for the interactions including
baryon resonances up to spin-5/2, and the explicit formulas relating the
coupling constants and the partial decay amplitudes were obtained.
With the predictions of the quark model, we then calculate the cross
sections and beam asymmetry of the reactions of $\gamma p \to K^+
\Sigma^0(1385)$ and $\gamma n \to K^+ \Sigma^-(1385)$, and the results are
compared to the recent data measured by the CLAS and LEPS
Collaborations.
Our results for the cross sections manifestly indicate a nontrivial role
of baryon resonances which are not-well-established or missing in the
list of the Particle Data Group.
However, the difficulty in explaining the measured beam asymmetry data
for the $\gamma n \to K^+ \Sigma^-(1385)$ reaction shows that much more
detailed analyses are required to understand the
production mechanisms of $\Sigma(1385)$.

%%%%%%%%%%%%%%%%%%%%%%%%%%%%%%%%%%%%%%%%%%%%%%%%
%% BACKMATTER
%%%%%%%%%%%%%%%%%%%%%%%%%%%%%%%%%%%%%%%%%%%%%%%%

\begin{theacknowledgments}

We are grateful to L. Guo, K. Hicks, C. M. Ko, K. Nakayama, and 
T.-S. H. Lee for fruitful discussions.
This work was supported by Kyungpook National University Research Fund, 2010.

\end{theacknowledgments}

\end{document}